\documentstyle[epsfig]{qcdparis}
\begin{document}
\pagestyle{plain}

%
%

\topmargin 0cm
\newcommand{\cernnr}{95-121}
\newcommand{\vdate}{May 1995}

\thispagestyle{empty}

\renewcommand{\thefootnote}{\fnsymbol{footnote}}
\setcounter{footnote}{0}

\hfill
\hspace*{\fill}
\begin{minipage}[t]{5cm}
\hfill CERN-TH/\cernnr
\end{minipage}

\vspace{0.5cm}
\vspace{1cm}

\begin{center}

{\Large\bf
   {Jet Physics in Deeply Inelastic Scattering at HERA}$\;^\ast${\it\fnm{7}
      {\em Talk presented
       at the Workshop on Deeply Inelastic
       Scattering and QCD, Paris, April 1995.
       To appear in the proceedings.
      }\\
   }
}

\vspace{1cm}
\vspace{0.5cm}


\vspace{0.1cm}

{\bf Dirk Graudenz}$\;^\sharp${\it\fnm{8}{{\em Electronic
mail address: Dirk.Graudenz\char64{}cern.ch.
}} \\
\vspace{0.1cm}
Theoretical Physics Division, CERN\\
CH--1211 Geneva 23\\
}

\vspace{2.0cm}

\begin{minipage}{14.0cm}

\begin{center}
{\bf Abstract}
\end{center}

\bigskip

We give an overview of jet physics in deeply inelastic scattering at HERA.
The problem of jet definitions, the scale dependence of jet cross sections
and some applications are discussed.

\vspace{1cm}
\begin{center}
{\bf R\'{e}sum\'{e}}
\end{center}

\bigskip

Nous donnons une vue d'ensemble de la physique des jets
en diffusion profond\'{e}ment in\'{e}lastique \`{a} HERA.
Le probl\`{e}me de la d\'{e}finition des jets, la d\'{e}pendance
d'\'{e}chelle des sections efficaces de production et quelques applications
sont discut\'{e}s.

\end{minipage}

\end{center}

\vfill
\noindent
\begin{minipage}[t]{5cm}
CERN-TH/\cernnr\\
\vdate
\end{minipage}
\vspace{1cm}

\noindent
\rule{80mm}{0.2mm}
\renewcommand{\thefootnote}{\arabic{footnote}}
\setcounter{footnote}{0}

\newpage

%
%

\newcommand{\epem}{$\mbox{e}^+ \mbox{e}^-$}
\newcommand{\reff}[1]{(\ref{#1})}
\newcommand{\nonu}{\nonumber\\}
\newcommand{\porder}[1]{\mbox{${\cal O}(#1)$}}
\newcommand{\yqi}{\mbox{$y_{qi}$}}
\newcommand{\yqg}{\mbox{$y_{qg}$}}
\newcommand{\yig}{\mbox{$y_{ig}$}}
\newcommand{\yqqb}{\mbox{$y_{q{\overline q}}$}}
\newcommand{\yqbi}{\mbox{$y_{{\overline q} i}$}}
\newcommand{\qbar}{\mbox{${\overline q}$}}
\newcommand{\iqbar}{\mbox{${\scriptstyle\overline q}$}}
\newcommand{\dilog}{{{\cal L}_2}}
\newcommand{\spence}{{\cal S}}
\newcommand{\lln}{l}
\newcommand{\cnumber}[0]{\mbox{\boldmath$C$}}
\newcommand{\rnumber}[0]{\mbox{\boldmath$R$}}
\newcommand{\nnumber}[0]{\mbox{\boldmath$N$}}
\newcommand{\munit}[0]{\mbox{\bf 1}}
\newcommand{\GeV}{\mbox{GeV}}
\newcommand{\SH}{{S_{\!\scriptscriptstyle H}}}
\newcommand{\XB}{{x_{\!\scriptscriptstyle B}}}
\newcommand{\grkl}{{\raisebox{-0.2cm}{$>$} \atop \raisebox{0.2cm}{$<$}}}

\title{Jet Physics in Deeply Inelastic Scattering at HERA$\,^\ast$}

\author{
Dirk Graudenz$\,^\sharp$}

\affil{CERN\\
--Theory Division--\\
CH-1211 Geneva 23}

\abstract{
We give an overview of jet physics in deeply inelastic scattering at HERA.
The problem of jet definitions, the scale dependence of jet cross sections
and some applications are discussed.
}

\resume{
Nous donnons une vue d'ensemble de la physique des jets
en diffusion profond\'{e}ment in\'{e}lastique \`{a} HERA.
Le probl\`{e}me de la d\'{e}finition des jets, la d\'{e}pendance
d'\'{e}chelle des sections efficaces de production et quelques applications
sont discut\'{e}s.
}

\twocolumn[\maketitle]
\fnm{7}{Talk given in the
Working Group Session on Hadronic Final States
at the Workshop on Deeply Inelastic Scattering and QCD,
Paris, April 1995.}
\fnm{8}{Electronic mail address:
Dirk.Graudenz@cern.ch.}

\renewcommand{\thefootnote}{\arabic{footnote}}
\setcounter{footnote}{0}

\section{Introduction}
The physics of jets and hadronic final states in deeply inelastic scattering
is a promising subject at HERA. Experimental results for jet production
have been reported by both the H1 \cite{1} and ZEUS \cite{2,3}
collaborations.
Owing to the large accessible range in
$Q^2$ and $\XB$, two important quantities (among others) can be measured:
the running coupling constant $\alpha_s(\mu_r^2)$ and the gluon density
$f_g(\xi,\mu_f^2)$$^{\mbox{\scriptsize\footnotemark}}$\fnm{\value{footnote}}{
$\xi$ is the momentum fraction carried by the incident parton,
and $\mu_r$ and $\mu_f$ are the renormalization and factorization scales,
respectively.}. The classification of hadronic final states according to the
number of jets and the idea that experimentally observed jets may be
identified with specific parton configurations defined by
jet algorithms
allow for a direct comparison of experimental data with theoretical
predictions. Particularly interesting
are (2+1) jet
events$^{\mbox{\scriptsize\footnotemark}}$\fnm{\value{footnote}}{i.e.\
events consisting of two jets in the
current region and one jet in the target fragmentation region.}, because
on the parton level, in perturbative QCD, the lowest-order process
for these events is of
\porder{\alpha_s}, and because the lowest-order diagram where the gluon density
comes in is of the type that gives rise to this particular final state.

Jet definitions should fulfil the following three criteria:

\noindent
(i) the definition should be given in terms of experimentally observable
quantities,

\noindent
(ii) it should be applicable in theoretical calculations
in the framework of perturbative QCD (i.e.\ it has to be infrared-safe
and should be easily formulated in terms of Lorentz-invariant quantities),
and

\noindent
(iii) it should be well-suited for the process under consideration,
e.g.\ adapted to the experimental situation.
\noindent
Algorithms which are presently used are of the following types:

\noindent
(a) Sterman--Weinberg-type algorithms based on energy and angle cuts
    \cite{4},

\noindent
(b) algorithms based on cones in pseudorapidity and azimuthal angle
    (for an overview, see \cite{5}),

\noindent
(c) cluster algorithms of the JADE-type \cite{6},

\noindent
(d) $k_T$-type algorithms \cite{7,8}.

\noindent
Algorithms from (a) and (b) are suited for \epem{} and pp colliders,
because they are adapted to specific frames of reference.
In deeply inelastic ep scattering, the CM frame of the QCD subprocess
is neither identical with the laboratory system nor related to this
frame by a boost along the beam axis alone. A Lorentz-invariant jet definition
scheme is therefore preferable, based on algorithms of type (c) and (d).
Presently available next-to-leading order (NLO) calculations are based
on modifications of the JADE algorithm. $k_T$-type algorithms are
constructed such that even the finite parts of the jet structure functions
factorize in the same way as totally inclusive structure functions like
$F_2$ do. This property allows for a resummation of terms
$\sim \log E_T^2/Q^2$ and $\sim \log y_{cut}$
\cite{7}. No theoretical NLO
calculations for jet cross sections are yet
available for $k_T$-type algorithms,
so we return to the discussion of the JADE-type algorithms.

Compared to jet physics in \epem{} annihilation, there is an additional
complication in ep scattering. One of the incoming particles is strongly
interacting and thus its remnants give rise to an additional jet.
In the perturbative picture, this is related to partons emitted
{}from the incident parton causing an infrared singularity in the collinear
phase-space region (see Fig.~\ref{fig1}, $p_2$ collinear to $p_r$).

\ffig{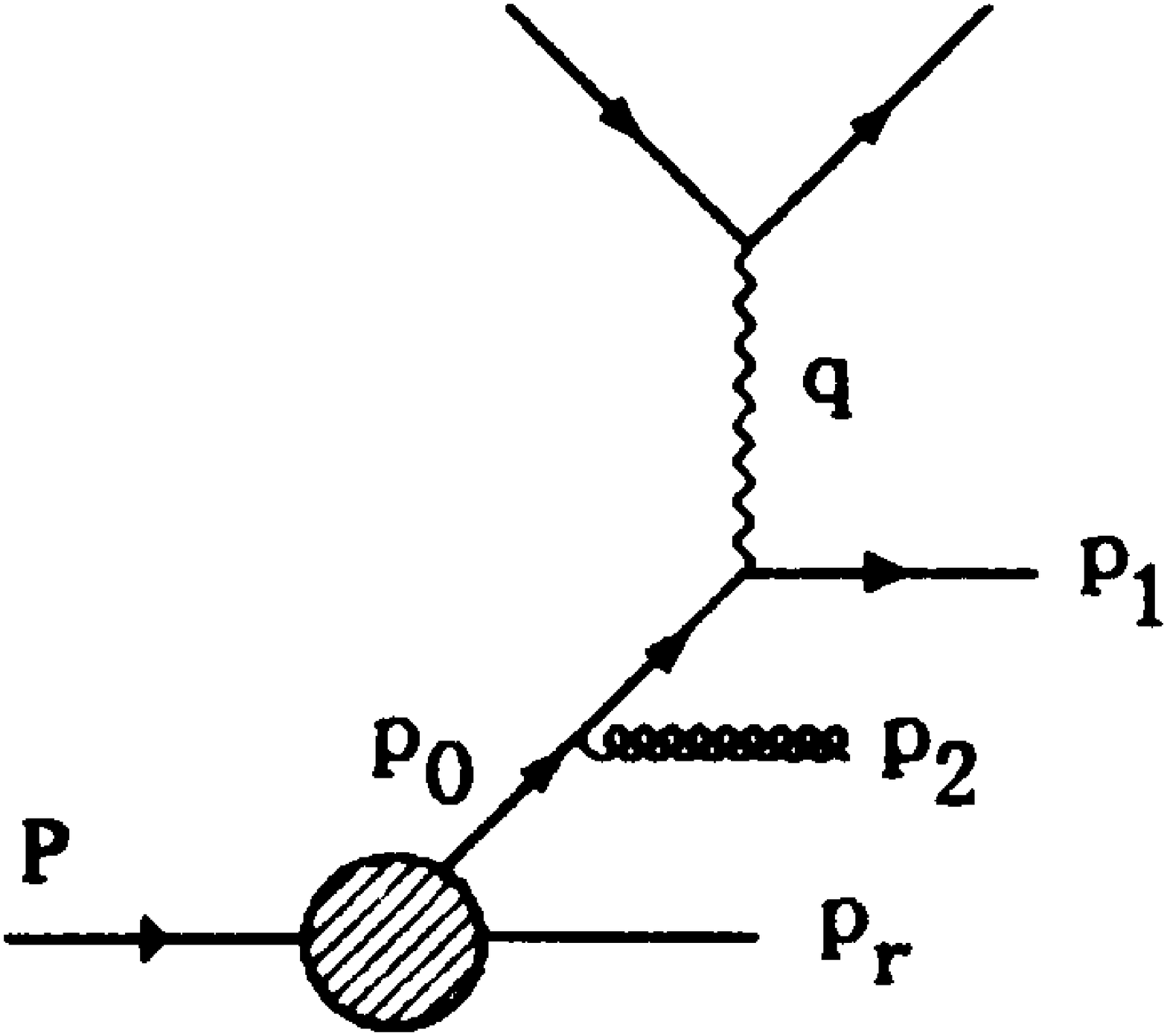}{40mm}{\em Feynman diagram giving rise to an
initial state singularity}{fig1}

The factorization properties of QCD take care of this singularity:
all collinear singularities from the initial state can be absorbed
in a process-independent way in universal parton distribution
functions. A consequence for jet physics is that a resolution criterion
must be given which specifies when a parton emitted from the incident parton
classifies as an additional jet: the proton remnant has to be included in
the jet definition.
A complication in the case of a collider experiment is that most of the remnant
jet simply disappears in the beam pipe without being seen by the detector.
The modified JADE (mJADE) algorithm is defined in the following way
\cite{9}:

\noindent
(1) define a {\it precluster} of longitudinal momentum $p_r$
given by the missing longitudinal momentum of the event,

\noindent
(2) apply the JADE cluster algorithm to the set of momenta
$\{p_1,\ldots,p_n,p_r\}$, where $p_1,\ldots,p_n$ are the momenta
of the visible hadrons in the detector. The resolution criterion
is $s_{ij}=2p_ip_j>cM^2$.
Here $M^2$ is a mass scale and
$c$ is the resolution parameter ($c\approx0.02$).

\noindent
In the case of a theoretical calculation, $p_r$ is directly given by
the momentum fraction of the proton not carried by the incident parton,
and $p_1,\ldots,p_n$ are the momenta of the partons in the final state.
In the following, we choose $W^2$, the squared total hadronic energy,
as the mass scale $M^2$, since the proton remnant is included in the
jet definition.

Based on a specific jet definition scheme, QCD corrections to the
leading order (LO) processes can be calculated. An overview will be given
in the next section. In Section~\ref{applications} we describe two
applications: The measurement of the running strong coupling constant
$\alpha_s(\mu_r^2)$ and the determination of the gluon density
$f_g(\xi,\mu_f^2)$ via jet rates.
The paper closes with a summary and conclusions.

\section{QCD corrections}
In the last few years, QCD corrections in NLO for jet production cross
sections have been calculated. The case of (1+1) jets is treated in detail
in \cite{10}. The NLO corrections to (2+1) jet production for the
dominant transverse photon helicity have been calculated in
\cite{11,12,13}. The remaining helicity cross sections
can be found in \cite{14}. The (3+1) jet cross section on the Born level
has been determined in \cite{15}, and even the (4+1) jet cross section
is known \cite{16}.
There are presently two programs available which incorporate NLO corrections:
DISJET \cite{17} (transverse and longitudinal cross sections, no arbitrary
acceptance cuts possible) and PROJET \cite{18} (all helicity cross sections
included, an event record allows for arbitrary acceptance cuts).
The numerical results presented in the following are based on PROJET.

\ffig{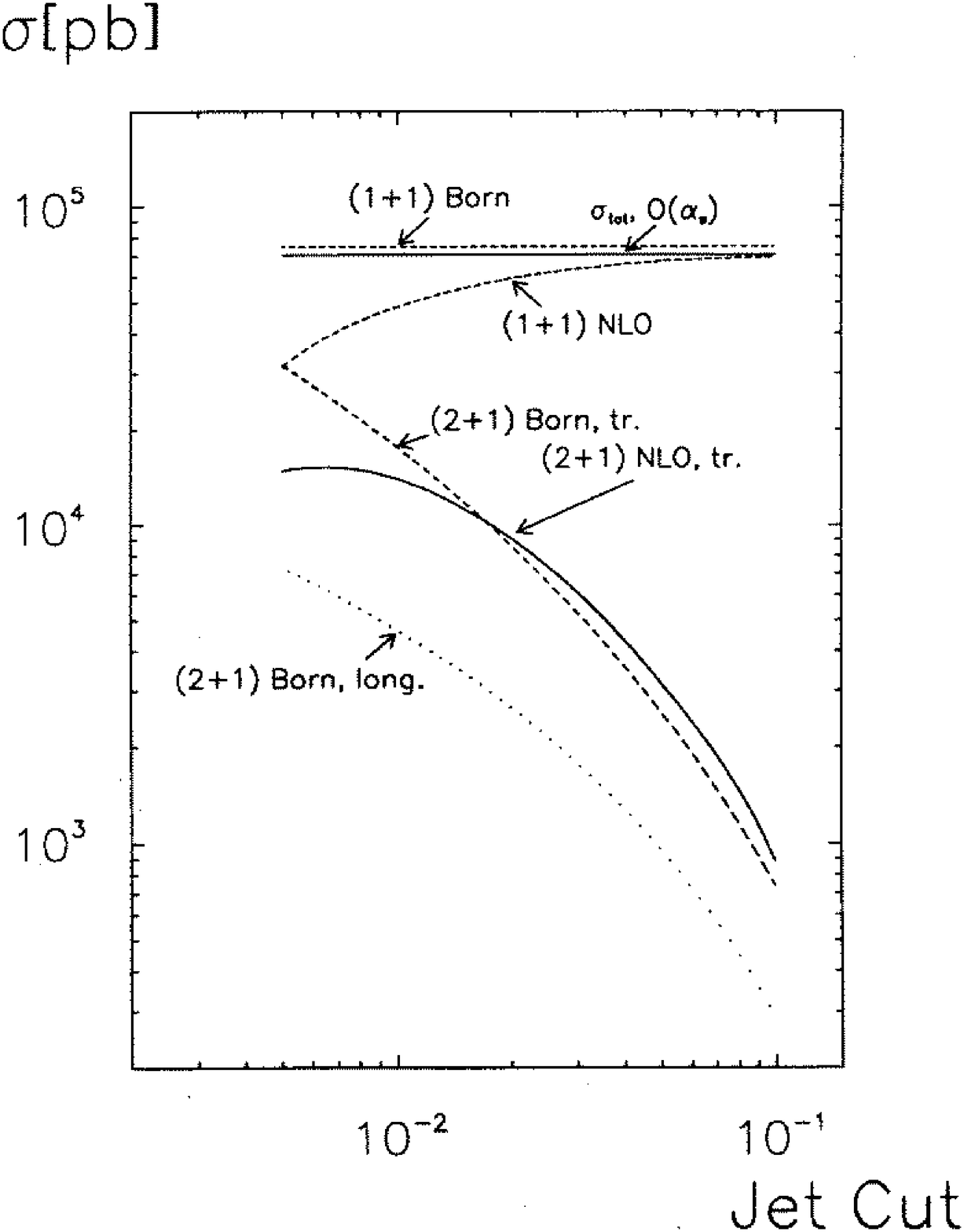}{80mm}{\em Jet cut dependence; ``tr.'' and ``long.''
stand for transverse and longitudinal contributions,
respectively}{fig2}

Figure~\ref{fig2} shows the dependence of the jet cross sections in LO and NLO
on the jet cut $c$. The kinematical parameters are
$E_{CM}=295\,\GeV$, $0.001<\XB<1$, $10\,\GeV<W<295\,\GeV$,
$3.16\,\GeV<Q<10\,\GeV$.
The parton density is MRS set D$^-$ \cite{19}.
The QCD corrections are moderate as long as $c>0.01$.

Fixed order perturbation theory introduces a scale dependence on the
renormalization scale $\mu_r$ and factorization scale $\mu_f$.
The NLO cross section for e.g.\ the (2+1) jet final state can be written
formally in the form
\begin{eqnarray}
\sigma^{NLO}(\mu_r^2,\mu_f^2)=\int\mbox{d}\xi \, f(\xi,\mu_f^2)
\,\big\{
\,\alpha_s(\mu_r^2)\,T_{Born}(\xi)\\ \nonumber
+\,\left(\alpha_s(\mu_r^2)\right)^2\left(
\,T_{virt.}(\xi,\mu_r^2)\,+\,T_{real}(\xi,\mu_f^2)\,
\right)
\big\}.
\end{eqnarray}
The variation of $\sigma^{NLO}(\mu_r^2,\mu_f^2)$
with $\mu_r^2$ and $\mu_f^2$ is of ${\cal O}(\alpha_s^3)$. It is thus
expected that the scale dependence of the NLO result is smaller than that
of the LO result.

\ffig{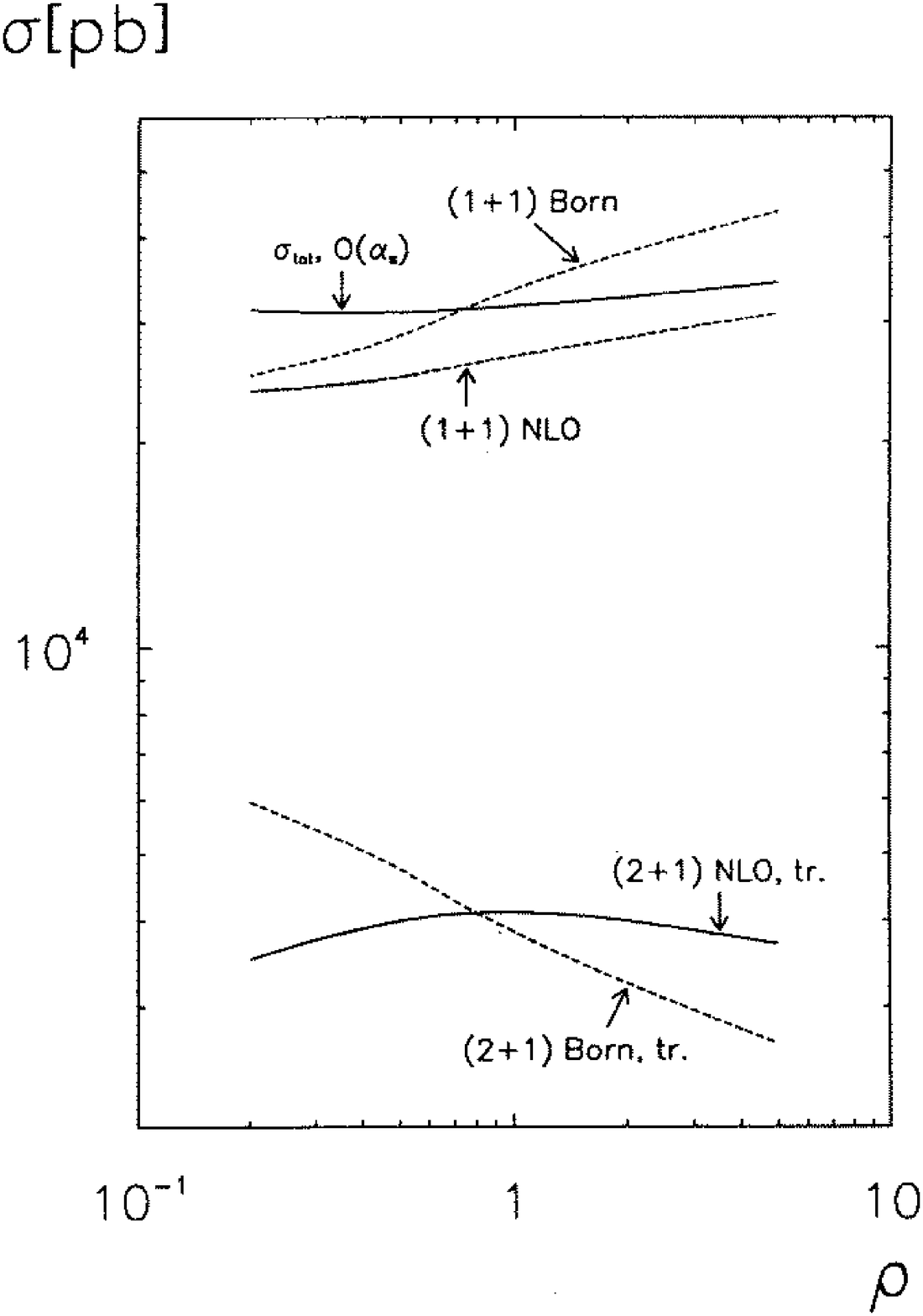}{80mm}{\em Scale dependence of jet cross sections}{fig3}

Figure~\ref{fig3} shows this dependence for
$\mu_r=\rho Q$, $\mu_f=\rho Q$. The parameters are
$E_{CM}=295\,\GeV$, $5\,\GeV<Q<100\,\GeV$, $10\,\GeV<W<295\,\GeV$, $c=0.02$.
The scale dependence of the NLO result is clearly reduced.

\ffig{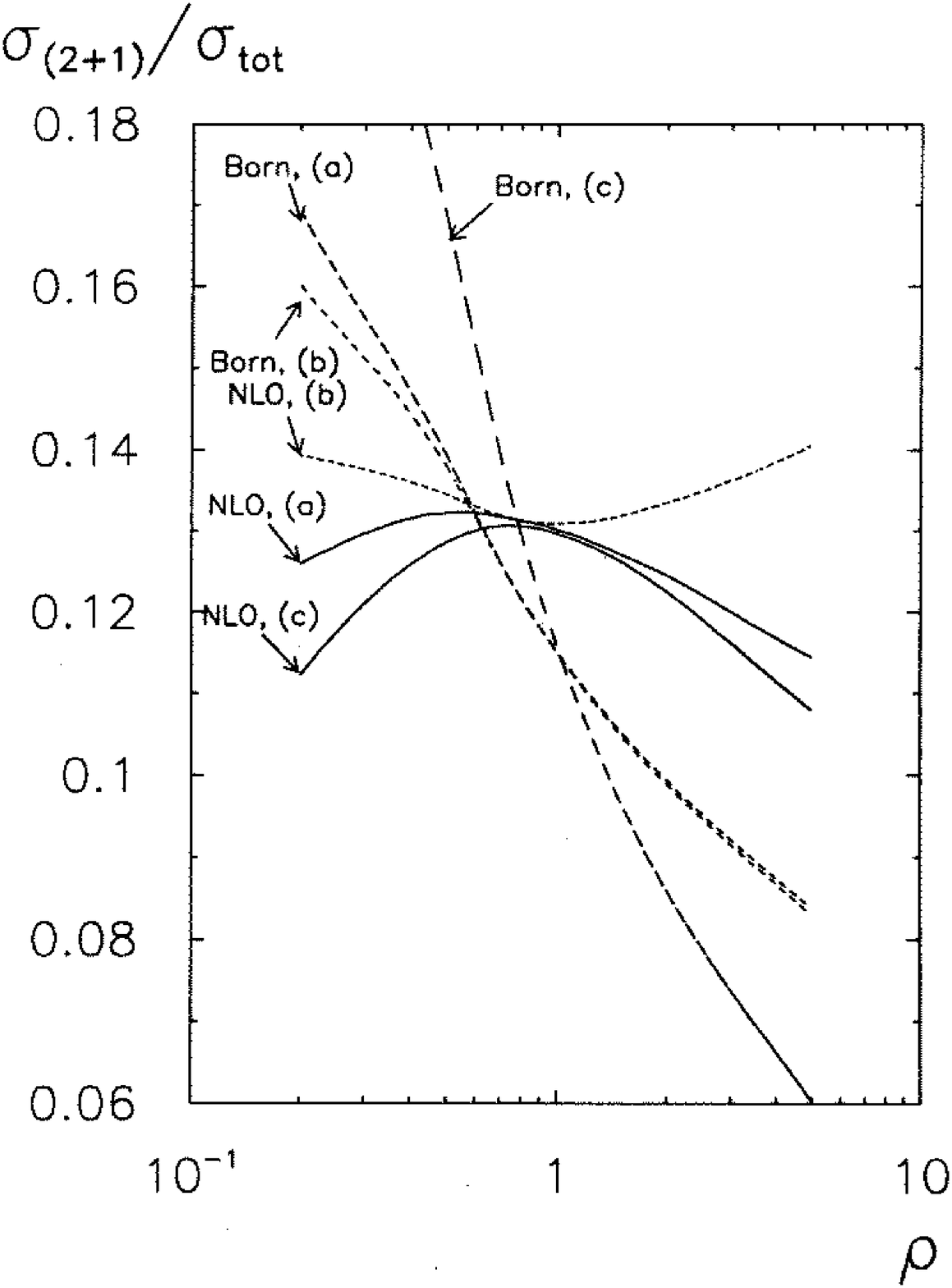}{80mm}{\em Scale dependence of the (2+1) jet rate}{fig4}

The effect on the (2+1) jet rate
$R_{2+1}=\sigma_{2+1}/\sigma_{tot}$ is displayed
in Fig.~\ref{fig4}, where (a), (b) and (c) stand for
$\mu_r=\rho Q$, $\mu_f=Q$;
$\mu_r=Q$, $\mu_f=\rho Q$;
and $\mu_r=\rho Q$, $\mu_f=\rho Q$,
respectively.

\section{Applications}
\label{applications}
We briefly discuss two interesting applications, the measurement of
the strong coupling constant and the determination of the gluon density.

In LO, the (2+1) jet rate $R_{2+1}$ is proportional to the strong coupling
constant $\alpha_s$. This makes $R_{2+1}$ an interesting observable, if
the hadronization corrections can be controlled.
In a recent paper, the H1 collaboration has quoted a value
$\alpha_s(M_Z^2)=0.123\pm0.018$ from $R_{2+1}$ \cite{20},
consistent with the current
world average.
One of the major uncertainties of this result is the systematic error
{}from the correction factors from the hadron level to the parton level.
More refined jet algorithms which are better adapted to the experimental
situation may improve the situation.

In the analysis, it turned out that the forward direction
is not yet well understood. Cuts on the polar angle and transverse momentum
were imposed in the H1 analysis
in order to restrict the phase space variables to a region where the
application of fixed order matrix elements is justified \cite{21}.
Cuts on the (2+1) jet variable $z$ as used by the ZEUS collaboration
in their analysis of jet cross sections
have a similar effect \cite{3}.

Another challenge is a direct determination of the gluon density
via jet rates. It has been pointed out \cite{22} that
the gluon density $f_g(\xi,\mu_f^2)$
is not yet well constrained for $0.01<\xi<0.1$. Precisely
in this region jet cross sections are sensitive to the parton densities.
An experimental analysis is therefore worth while.
The gluon density $f_g$ can be reconstructed from
\begin{equation}
R_{2+1}^{exp}=\frac{
f_q\otimes\sigma_{2+1,q}^{th}+f_g\otimes\sigma_{2+1,g}^{th}
}
{
f_q\otimes\sigma_{tot,q}^{th}+f_g\otimes\sigma_{tot,g}^{th}
}
\end{equation}
if the quark densities $f_q$ are known.
Recently, first experimental results in LO have been obtained \cite{23}.
A feasibility
study in LO based on an event generator is available \cite{24}.
The applied method is based on a direct reconstruction of the momentum fraction
$\xi$ of the incident parton from the final state kinematics. Unfortunately,
this simple and straightforward
method breaks down in NLO, because then $\xi$ is no longer
an observable, due to the redefinition of the parton densities and
corresponding finite subtractions. A method
which is applicable in NLO
and based on the Mellin transform
is presently under study \cite{25}.

\section{Summary and Conclusions}
We have given a brief overview of jet physics in deeply inelastic
electron--proton scattering with some emphasis on next-to-leading order
QCD corrections. The corrections stabilize the theoretical predictions
with respect to scale variations and allow for an experimental determination
of scale-dependent quantities.

The NLO predictions have already been used for a
determination of
the running strong coupling constant $\alpha_s(\mu_r^2)$ at HERA.
A measurement of the
gluon
density $f_g(\xi,\mu_f^2)$ in NLO for $\xi>0.01$ by means of jet rates
seems to be feasible as well.

It would certainly be desirable to have NLO predictions for jet cross sections
based on other jet definition schemes, such as the $k_T$-scheme and
cone algorithms, as well. As can easily be seen,
the mJADE algorithm restricts the accessible
range in the momentum fraction $\xi$ of the incident parton to values larger
than the jet cut $c$, if the mass scale used for the jet definition is $W^2$.
In order to be sensible to the gluon density at small $\xi$,
other jet definitions have to be used.
Moreover, event-shape variables for ep scattering
should be
considered as well.

{}From a physics point of view, the forward (i.e.\ proton) direction deserves
further study. Jet production in this region is not yet well understood,
and it is unclear whether fixed order matrix elements can describe the
situation at all.
A study with event generators shows that the problem probably stems from
the emission of partons from the incident parton, modelled by an initial
state parton shower. Moreover, the fragmentation
of the target remnant jet is poorly understood and contributes to the problem.

\begin{center}
{\large\bf Acknowledgements}
\end{center}
It is a pleasure to thank the conveners of the
working group on Hadronic Final States and the organizing committee.
\vspace{2cm}
\Bibliography{XXX}
\newcommand{\bibitema}[1]{\bibitem[#1]{#1}}
%


%
\bibitema{1}
      H1 Collaboration, Z.\ Phys.\ \underline{C61} (1994) 59.

\bibitema{2}
      ZEUS Collaboration, Phys.\ Lett.\ \underline{B306} (1993) 158.

\bibitema{3}
      ZEUS Collaboration, preprint DESY 95-016 (February 1995),
      to be published in Z.\ Phys.\ \underline{C}.

\bibitema{4}
      G.~Sterman, S.~Weinberg,
      Phys.\ Rev.\ Lett.\ \underline{39} (1977) 1436.

\bibitema{5}
   J.E.~Huth, N.~Wainer, K.~Meier, N.~Hadley, F.~Aversa,
   M.~Greco, P.~Chiappetta, J.P.~Guillet, S.~Ellis,
   Z.~Kunszt, D.~Soper, preprint FERMILAB-CONF-90-249-E,
   December 1990,
   published in Snowmass Summer Study 1990.

\bibitema{6}
      JADE Collaboration, W.~Bartel et al.,
      Z.\ Phys.\ \underline{C33} (1986) 23.

\bibitema{7}
      S.~Catani, Y.L.~Dokshitzer, B.R.~Webber,
      Phys.\ Lett.\ \underline{B285} (1992) 291.

\bibitema{8}
      B.~Webber, J.\ Phys.\ \underline{G19} (1993) 1567.

\bibitema{9}
      D.~Graudenz, N.~Magnussen,
      in: Proceedings of the HERA Workshop 1991,
      DESY (eds.\ W.~Buchm\"uller, G.~Ingelman).

\bibitema{10}
      J.G.~K\"orner, E.~Mirkes, G.A.~Schuler,
      Int.\ J.\ Mod.\ Phys.\ \underline{A4} (1989) 1781.

\bibitema{11}
      D.~Graudenz, {\it Der Drei-Jet-Wirkungs\-quer\-schnitt zur Ordnung
      $\alpha_s^2$ in der tief-in\-elast\-isch\-en
      El\-ek\-tron-Proton-Streuung},
      Ph.D.\ thesis,
      DESY-T-90-01 (September 1990).

\bibitema{12}
      D.~Graudenz,
      Phys.\ Lett.\ \underline{B256} (1991).

\bibitema{13}
      D.~Graudenz, Phys.\ Rev.\ \underline{D49} (1994) 3291.

\bibitema{14}
      T.~Brodkorb, J.G.~K\"orner,
      Z.\ Phys.\ \underline{C54} (1992) 519.

\bibitema{15}
      T.~Brodkorb, J.G.~K\"orner, E.~Mirkes, G.A.~Schuler,
      Z.\ Phys.\ \underline{C44} (1989) 415.

\bibitema{16}
      A.~Dabelstein, preprint DESY-90-148 (November 1990).

\bibitema{17}
      T.~Brodkorb, E.~Mirkes, preprint MAD-PH-821 (Madison University, 1994).

\bibitema{18}
      D.~Graudenz, PROJET 4.13 manual, preprint CERN-TH.7420/94
      (November 1994).

\bibitema{19}
      A.~Martin, R.~Roberts, J.~Stirling,
      Phys.\ Rev.\ \underline{D47} (1993) 867.

\bibitema{20}
      H1 Collaboration, Phys.\ Lett.\ \underline{B346} (1995) 415.

\bibitema{21}
      R.~Nisius, {\it Measurement of the strong coupling constant $\alpha_s$
      from jet rates in deep inelastic scattering},
      Ph.D.\ thesis, PITHA-94/21 (Aachen University, 1994).

\bibitema{22}
      J.~Stirling, these proceedings.

\bibitema{23}
      H1 Collaboration, preprint DESY-95-086 (1995).

\bibitema{24}
      A.~Edin, preprint TSL/ISV-93-0087 (Uppsala University, 1993).

\bibitema{25}
      D.~Graudenz, M.~Hampel, A.~Vogt, Ch.~Berger,
      in preparation.

\end{thebibliography}
\end{document}